\documentclass{aa}  

\usepackage{graphicx}
\usepackage{txfonts}
\usepackage{color}

\newcommand{\inte}{\textsl{INTEGRAL}}

\def\ergcms{{\rm erg\,cm^{-2}\,s^{-1}}}

\begin{document}

   \title{Rapid spectral transition of the black hole binary V404~Cyg}


   \author{J.~J.~E. Kajava
          \inst{1,2,3}
          \and
          C. S\'anchez-Fern\'andez\inst{2}
          \and
          J. Alfonso-Garz\'on\inst{1}
          \and
          S.~E. Motta\inst{4}
          \and
          A. Veledina\inst{5, 6, 7}
          }

   \institute{Centro de Astrobiolog\'{\i}a -- Departamento de Astrof\'{\i}sica (CSIC-INTA), Camino Bajo del Castillo s/n, E-28692 Villanueva de la Ca\~nada, Spain\\
    \email{jkajava@cab.inta-csic.es}
    \and
    European Space Astronomy Centre (ESA/ESAC), Science Operations Department, E-28691, Villanueva de la Ca\~{n}ada, Madrid, Spain
    \and
    Finnish Centre for Astronomy with ESO (FINCA), FI-20014 University of Turku, Finland
    \and
    University of Oxford, Department of Physics, Astrophysics, Denys Wilkinson Building, Keble Road, Oxford OX1 3RH, UK     
    \and
    Department of Physics and Astronomy, FI-20014 University of Turku, Finland
    \and
    Nordita, KTH Royal Institute of Technology and Stockholm University, Roslagstullsbacken 23, SE-10691 Stockholm, Sweden
    \and
    Space Research Institute of the Russian Academy of Sciences, Profsoyuznaya Str. 84/32, 117997 Moscow,  Russia
     \\
             }

   \date{Received Nov 27, 2019; accepted Jan 10, 2020}

\abstract{During the June 2015 outburst of the black hole binary V404 Cyg, rapid changes in the X-ray brightness and spectra were common.
The \textit{INTEGRAL} monitoring campaign detected spectacular Eddington-limited X-ray flares, but also rapid variations at much lower flux levels.
On 2015 June 21 at 20~h~50~min, the 3--10~keV JEM-X data as well as simultaneous optical data started to display a gradual brightening from one of these low-flux states. 
This was followed 15 min later by an order-of-magnitude increase of flux in the 20--40~keV IBIS/ISGRI light curve in just 15 s.
The best-fitting model for both the pre- and post-transition spectra required a Compton-thick partially covering absorber. 
The absorber parameters remained constant, but the spectral slope varied significantly during the event, with the photon index decreasing from $\Gamma \approx 3.7$ to $\Gamma \approx 2.3$.
We propose that the rapid 20--40 keV flux increase was either caused by a spectral state transition that was hidden from our direct view, or that there was a sudden reduction in the amount of Compton down-scattering of the primary X-ray emission in the disk outflow.} 
 


   \maketitle
%

\section{Introduction}

Black hole binaries (BHB) exhibit various X-ray spectral states when they accrete gas during transient outburst episodes (for review, see, e.g., \citealt{DGK07,Belloni2016}).
The two main spectral states are the hard and the soft state, historically also called low-hard state (LHS) and high-soft states (HSS), respectively (see, e.g., \citealt{TGK72}).
In the LHS, the X-ray spectra are described by a hard power law of $\Gamma \sim 1.7$ with a high-energy cutoff at about 30--100~keV that is attributed to thermal Comptonization in a ``corona'' or a hot medium near the BH \citep{ST80}.
In the HSS the X-ray spectrum is instead dominated by a $\sim$1~keV thermal disk component, with a weak high-energy tail that becomes significant above $\sim$20~keV.
Sometimes the spectra show emission line(s) from neutral and/or ionized iron at 6.4--6.97~keV and a Compton reflection hump at higher energies (see, e.g., \citealt{Reynolds2014}), indicating that the Comptonized emission is reflected off a cool accretion disk that can be truncated at tens to hundreds of gravitational radii from the BH.
More recent spectro-timing state classifications have introduced intermediate states between the two main ones: the hard-intermediate and the soft-intermediate states (HIMS and SIMS, respectively, \citealt{HWvdK01, Belloni2016}). 
In HIMS and SIMS the energy spectra show properties of both the LHS and the HSS, observed to evolve smoothly during the hard-to-soft (or soft-to-hard) transitions, while the most striking changes are observed in the time domain (see, e.g., \citealt{Nespoli2003}).

During a transient X-ray outburst a source may go through all of these states in a specific order, which leads to the well-known spectral hysteresis behavior \citep{Miyamoto1995,HWvdK01,MC03}.
In addition to these common spectral states, some sources (such as GRO~J1655--40, GRS~1915+105, XTE~J1550--564, GX~339--4, and V404~Cyg) can visit the so-called very high state (VHS; also known as the ultra-luminous state, steep power-law state, or the anomalous state) at the highest luminosities, which is characterized by a soft power-law spectrum, with a photon index of $\Gamma\sim 2.5-3.0$ that can extend to MeV energies with no signs of a spectral cutoff \citep{TKK99,ZGP01,GD03,TKY12,Sanchez-Fernandez2017}.
Moreover, the BHBs Cyg~X--3 \citep{SZMc08,HZS09,KHMcC10} and GRO~J1655--40 \citep{UKW15} occasionally enter the ultra- or hypersoft states (USS), where the power-law emission can reach photon indices in the range of $\Gamma \sim 3-6$.

In this paper we focus on V404~Cyg, whose X-ray outbursts behavior is highly anomalous compared to the rest of the BH transients.
Its X-ray outbursts in 1989 \citep{T89,OvKV96,ZDS99}, in June 2015 \citep{RCBA15,RJB15,JWHH16} and December 2015 \citep{MCM17,KMS18} lasted only a few weeks, and the light curves are littered with bright flares that reach the Eddington limit and with brief periods where the X-ray flux decreased down to 0.1\%\  from the Eddington value.

X-ray spectral studies of V404~Cyg have shown that the flares are generated by two processes: 
The overall flux levels vary under the significant fluctuations of the mass accretion rate on hourly timescales, but occasionally, the variations are caused by the presence of Compton-thick absorbers, as inferred from the X-ray spectral modeling \citep{OvKV96,ZDS99,MKS17a, Sanchez-Fernandez2017, MKS17b, KMS18}.
During these epochs, the equivalent hydrogen column density can exceed $N_\textrm{H} \approx 1.5 \times 10^{24}\,\textrm{cm}^{-2}$, such that the Thomson optical depth through the absorber is above unity. 
This causes a large fraction of the light to either scatter away from our line of sight or to become absorbed, thus reducing the observed X-ray flux. 
This variable local absorption is due to an accretion disk wind, which apparently chops up longer X-ray flares into shorter ones as thick clouds or clumps pass in front of the central X-ray engine.
The disk winds also manifested themselves as P Cyg emission lines in X-ray data with high spectral resolution  \citep{King2015} and in optical spectra \citep{MDCMS16,CMM2019}.
The local absorption causes the hardness-intensity diagram (HID) to be dissimilar to other systems \citep{WMT17}, which further complicates comparisons to other BHBs.

The X-ray spectral analysis also shows that some epochs have no local absorption, in some epochs absorbing clouds change from Compton thick to Compton thin, and in some other epochs the X-ray flux evolves rapidly together with X-ray spectral hardness \citep{Sanchez-Fernandez2017, MKS17b}.
In this paper we concentrate on one of these spectral hardening events that occurred on 2015 June 21 (MJD~57194). 
It coincided with an order-of-magnitude increase of the hard X-ray flux from a relatively low level of $F_\textrm{20-80\,keV} \sim 2.5 \times 10^{-9}\,\ergcms$, which corresponds to roughly 0.2\%\  of the Eddington limit for V404~Cyg at an assumed distance of 2.39 kpc \citep{MJJD09} and to a black hole of 9 solar masses \citep{Khargharia2010}.

\section{Observations of V404~Cyg}

Soon after the first indications of renewed activity of V404~Cyg by the \textit{Niel Gehrels Swift} observatory \citep{Gehrels2004} on 2015 June 15, the INTErnational Gamma-Ray Astrophysics Laboratory (\inte; \citealt{W03}) initiated an extensive Target of Opportunity (ToO) campaign \citep{Kuulkers2015b, Kuulkers2015c}.
V404~Cyg was monitored almost continuously for almost four weeks, with only a few gaps in between due to perigee passages and other ToO targets.
In addition to the X-ray coverage with JEM-X \citep{LBJW03}, soft gamma-ray coverage with IBIS/ISGRI \citep{LLL03} and SPI (which we do not use here), the Optical Monitoring Camera (OMC) \citep{MHG03} provided simultaneous optical photometric observations in the \textit{V} band.
The available OMC data are complemented in this paper with public ground-based optical observations from the American Association of Variable Star Observers (AAVSO).

The JEM-X and ISGRI light curves and spectra were extracted using OSA~v.10.2 \citep{Courvoisier2003} following the standard reduction steps.
Light curves were extracted using 5s time bins in the JEM-X 3--10\,keV band and in the ISGRI 20--40, 40--80, and 80--200\,keV bands to study the flare profiles in detail and to accurately pinpoint the onset of the transition. 
We used the times derived from the inspection of the light curves to generate good time intervals (GTIs) that were used in the JEM-X and ISGRI spectral extraction. 
The JEM-X and ISGRI responses were logarithmically rebinned to obtain 23 and 28 channels, respectively.
The OMC light curves were also extracted with OSA v.10.2, as detailed in \citet{AGSF18}.

\section{Results}

\begin{figure*}
   \centering
  \includegraphics[width=0.95\textwidth]{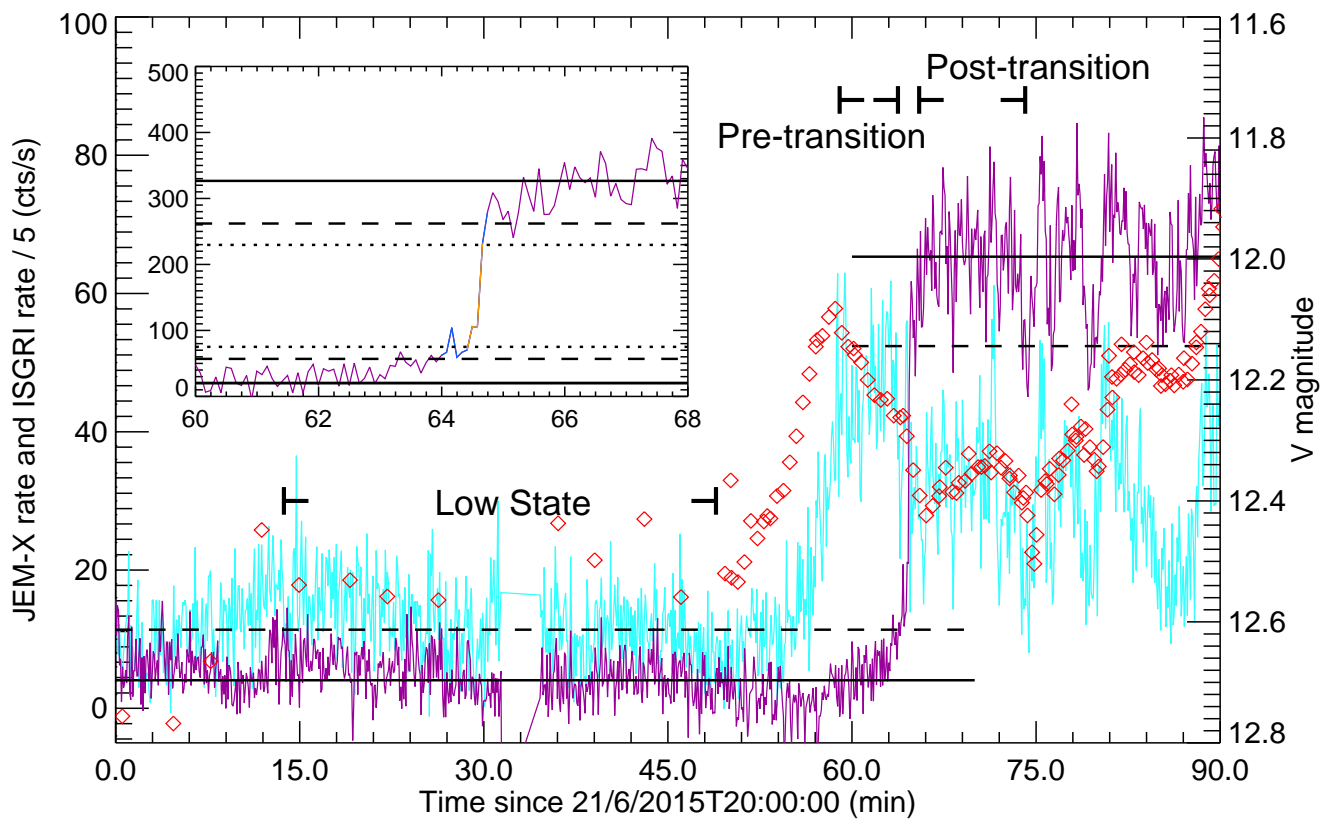}
   \caption{
   Light curves of the transition. The cyan line shows the JEM-X count rate (3--10 keV; left y-axis), the purple line shows the ISGRI count rate (20--40 keV; y-axis, count rate divided by 5 for better scaling), and the red diamonds show the optical \textit{V}-filter light curve (right axis). The solid and dashed lines show the low-state and the post-transition mean ISGRI count rate and the 2-$\sigma$ deviation from it, respectively. The inset shows a zoom-in of the transition, with blue and yellow lines indicating the deviation from 2-$\sigma$ and 3-$\sigma$ deviation (dotted line) from the low-state and post-transition mean count rates, respectively.
   } 
\label{fig:lightcurves}
\end{figure*}

The multiwavelength light curves that cover the transition are shown in Fig.~\ref{fig:lightcurves}.
The data show that the optical $V$-band and the soft X-ray brightness first begins to rise at around 20~h~54~min, while the hard X-ray flux remains constant (see \citealt{AGSF18}, for a more detailed cross-correlation analysis).
The \textit{V} -band and soft X-ray brightness hit a local maximum at the same time, which is then followed by a leveling of the JEM-X count rate and an immediate and gradual decrease of the optical flux.
At around 21~h~03~min, the ISGRI flux also begins to rise. The initial flux increase is gradual, but is then amplified at 21~h~04~min~30~s with an episode of rapid flux increase.
During the fast transition, the optical brightness drops by 0.2 magnitudes, and the JEM-X count rates drop similarly rapidly from about 45 counts per second (cps) to 30 cps.
The ISGRI flux jump from 75 cps to 230 cps occurs in 15~s (within the 3$\sigma$ limits of the low and high flux levels, see dotted lines in the inset), which can be considered to be the transition timescale.

\begin{figure}
   \centering
  \includegraphics[]{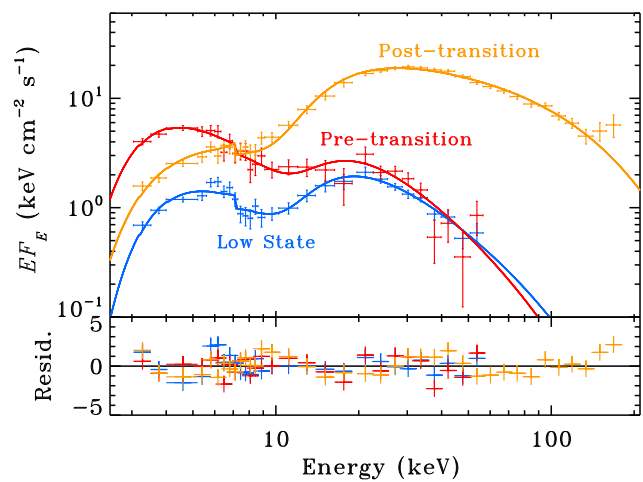}
   \caption{Rapid spectral changes seen during the transition.}
    \label{fig:transition}%
\end{figure}

To better understand the nature of this transition, we extracted three spectra from this event during the periods highlighted in Fig.~\ref{fig:lightcurves}.
The low-state spectrum was averaged over time before the optical flux rise had started.
The pre-transition spectrum was extracted from the time of maximum optical flux (at the constant
JEM-X flux maximum) until the time when the ISGRI flux started to rise.
The post-transition spectrum was averaged over the period after the ISGRI rise and before the secondary optical increase, that is, during the time when the ISGRI flux was rather stable.
The three spectra are shown in Fig.~\ref{fig:transition}.
We also extracted a spectrum from the interval between the low state and the pre-transition periods. 
This spectrum showed intermediate hardness and flux, but was naturally much noisier and was therefore not considered further in the paper.

The \textsc{xspec} model we used to describe the continuum is the thermal Comptonization model \textsc{nthcomp} \citep{ZJM96, ZDS99}.
The fitting parameters are the photon index, $\Gamma$, the electron temperature $T_\textrm{e}$ , and a normalization factor.
The seed photon temperature was fixed to 0.1~keV, except for one case, as outlined below.
The local and interstellar absorption was taken into account with the \textsc{tbnew} model (an updated version of the \citealt{Wilms2000} \textsc{tbabs} model).
It was required to have at least the interstellar value of $N_\textrm{H,\,gal} = 0.83\times 10^{22}$~cm$^{-2}$.
We also tested if instrument cross-normalizations between JEM-X and ISGRI were required, but they were found to be consistent with unity in the best-fitting model and thus were not used in the results reported below. 

We first fit the low state, pre-transition, and post-transition spectra with the simple one-absorber model \textsc{tbnew} $\times$ \textsc{nthcomp}, allowing the absorption column to vary freely.
The best-fitting parameters are shown in Table \ref{tab:nthComp}. 
The one-absorber model is clearly too simplistic.
The electron temperature can be constrained only in the post-transition spectrum with the highest signal-to-noise ratio, but to a very low value of $T_\textrm{e} \approx 13$~keV.
The model always fits the data poorly, with  $\chi^2/{\rm d.o.f.} \gtrsim 10$ in the worst case. 
We therefore added three more components to the model: the disk model \textsc{diskbb}, the reflection model \textsc{reflect,} and a Gaussian at 6.4~keV to mimic an iron line. 
The poor spectral resolution of JEM-X does not warrant the use of more sophisticated reflection models, such as the RELXILL model \citep{GDL2014} that was used by \citet{WMT17} to fit the NuSTAR data of V404 Cyg.
We refer to this case as the ``reflection model''. It fits the data much better and results in a more reasonable electron temperature of $T_\textrm{e} \approx 60$ keV for the post-transition spectrum (for comparison, see \citealt{Sanchez-Fernandez2017}).
However, only in the pre-transition spectrum is the model statistically acceptable, but in the low-state and the post-transition state, the reflection model can be rejected with high confidence ($\chi^2/{\rm d.o.f.} \gtrsim 2$ and null hypothesis probability $P_{\rm null} \sim 10^{-4}$).
In all cases the reflection fraction is very high, effectively suggesting that we only see the reflection of the source and not the source itself.
The addition of the Gaussian line improves the fit only in the low-state case, and the \textsc{diskbb} model improves the fit only in the pre-transition case. 
The \textsc{diskbb} model normalization, $K_\textrm{dbb}$, is strongly correlated with the column density, which results in large uncertainties in both quantities.
$K_\textrm{dbb}$ can still be used to obtain an order-of-magnitude estimate of the inner disk radius, giving $R_\textrm{dbb} = d_{10} \sqrt{K_\textrm{dbb}/\cos{i}} \approx 12$~km (where $d_{10} = 0.239$ is the distance in units of 10~kpc and $i=67\degr$ is assumed). 
This is significantly smaller than the innermost stable circular orbit (ISCO) $R_\textrm{ISCO} \approx 80$~km for a nonrotating 9~M$_{\sun}$ BH, but is consistent with a maximally spinning one (see, e.g., \citealt{Bardeen1972, Makishima2000}).
However, as shown below, we note that the data are much better modeled with a partially covering absorber without a disk component, which casts doubt on this order-of-magnitude estimate.
The local absorption column in this model evolves across the transition: from a nonabsorbed and moderately soft low-flux state spectrum with $\Gamma \approx 2.5$ into a modestly absorbed and softer, $\Gamma \approx 2.9$, pre-transition spectrum.
The post-transition spectrum is characterized by simultaneous disappearance of the disk component and the local absorber, together with a significant hardening of the spectrum to $\Gamma \approx 1.9$.

In the dual-absorber spectral model we added a partially covering absorber with \textsc{tbnew\_pcf}.
This model is statistically preferred, yielding $\chi^2_\textrm{red} \approx 1.0$ for all three spectra.
Neither Gaussian lines nor disk components are needed in this case.
Interestingly, the best-fitting parameters for the two absorbers do not change significantly.
In all three spectra the fully covering column is $N_\textrm{H} \approx 2\times 10^{23}$~cm$^{-2}$, while the partially covering column is Compton-thick with $N_\textrm{H,\,pcf} \approx 5\times 10^{24}$~cm$^{-2}$ with a partial covering fraction of $\textrm{PCF} \approx 0.9$ (see also \citealt{MKS17b}).
The only parameter that evolves significantly is the photon index, which changes from a value consistent with the hypersoft state $\Gamma \approx 3.7$ (pre-transition) to $\Gamma \approx 2.3$ (post-transition). 
We note that none of these considered models is able to explain the apparent excess in the post-transition spectrum near 200 keV, which could be from the same source component as that which causes the gamma-ray excess seen by \inte/SPI \citep{SDG16, JRR17}.
 
\begin{table*}
\centering
\caption{\label{tab:nthComp}Best-fitting parameters for the \textit{INTEGRAL} low-state, pre-, and post-transition spectra with the one-absorber, reflection, and dual-absorber models. 
The Galactic absorption column $N_{\rm H}$ is given in units of $10^{22}$~cm$^{-2}$ and is fixed to the interstellar value of $0.83$ (values in square brackets), unless letting it vary led to a statistically significant improvement of the fit.
The partial covering absorber column density $N_{\rm H,\,pcf}$ is also given in units of $10^{22}$~cm$^{-2}$.
The seed photon temperature was fixed to 0.1~keV, except in the pre-transition reflection model, where it was tied to the \textsc{diskbb} model temperature $T_{\rm dbb}$.
The $T_{\rm dbb}$ is given in units of keV.
The \textsc{diskbb} model normalization is $K_{\rm dbb} = (R_\textrm{dbb}$[km]/$d_{10})^2 \cos i$.
$T_{\rm e}$ is the electron temperature in units of keV, which can be constrained only in post-transition spectrum.
For other spectra, its values were fixed. 
The reflection is computed as ${\rm Refl}=\Omega/2\pi$, where $\Omega$ is the solid angle occupied by the reflecting medium.
The 20--80~keV band fluxes are not corrected for local or interstellar absorption, and they are given in units of $10^{-9}\,{\rm erg\,cm^{-2}\,s^{-1}}$.
} 
\begin{tabular}{@{}lcccccccccr}
\hline\hline
ID  & $N_{\rm H}$                    & $N_{\rm H,\,pcf}$    & PCF             & $T_{\rm dbb}$      & $K_{\rm dbb}$             & $\Gamma$                      & $T_{\rm e}$                 & Refl                        & $F_{20-80}$                   & $\chi^{2}/\textrm{d.o.f.}$ \\
\hline
\multicolumn{11}{c}{Low state} \\
\hline
  One absorber & $6.2_{-1.3}^{+1.4}$         & ...       & ...    & ... & ...  & $1.93_{-0.03}^{+0.03}$                & $[13.3]$         & ...        & $2.04_{-0.08}^{+0.07}$      &  108.8/25 \\
 Reflection & $[0.83]$         & ...       & ...    & ... & ...  & $2.51_{-0.05}^{+0.05}$                & $[60]$         & $13_{-3}^{+4}$        & $2.10_{-0.15}^{+0.04}$      &  57.0/24 \\
  Dual absorber & $26_{-3}^{+3}$          & $590_{-60}^{+60}$    & $0.90_{-0.02}^{+0.02}$       & ...  & ...  &  $3.37_{-0.13}^{+0.13}$                & $[62]$         & ...        & $2.13_{-0.09}^{+0.06}$      &  23.1/23 \\
\hline
\multicolumn{11}{c}{Pre-transition} \\
\hline
One absorber  & $3.1_{-1.4}^{+1.4}$         & ...       & ...    & ... & ...  & $2.40_{-0.05}^{+0.05}$                & $[13.3]$         & ...        & $2.21_{-0.15}^{+0.13}$      &  65.5/25 \\
 Reflection & $10_{-4}^{+5}$         & ...       & ...  & $1.08_{-0.15}^{+0.15}$   & $1000_{-500}^{+2000}$  & $2.91_{-0.11}^{+0.11}$                & $[60]$         & $-1$        & $2.45_{-0.7}^{+0.03}$      &  27.7/23 \\
Dual absorber &  $19_{-3}^{+3}$          & $680_{-110}^{+140}$    & $0.87_{-0.05}^{+0.04}$       & ...  & ...  &  $3.7_{-0.2}^{+0.2}$                & $[62]$         & ...        & $2.48_{-0.2}^{+0.13}$      &  24.6/23 \\
\hline
\multicolumn{11}{c}{Post-transition} \\ 
\hline
 One absorber & $11_{-2}^{+2}$         & ...       & ...    & ... & ...  & $1.40_{-0.02}^{+0.02}$                & $13.3_{-0.3}^{+0.3}$         & ...        & $34.5_{-0.5}^{+0.3}$      &  395.8/34 \\
Reflection & $[0.83]$         & ...       & ...  & ...   & ...  & $1.87_{-0.03}^{+0.03}$                & $60_{-9}^{+13}$         & $25_{-5}^{+7}$       & $35.1_{-0.4}^{+0.3}$      &  73.4/34 \\
Dual absorber  & $17_{-2}^{+2}$          & $490_{-40}^{+40}$    & $0.894_{-0.012}^{+0.011}$       & ...  & ...  &  $2.30_{-0.06}^{+0.07}$                & $62_{-14}^{+30}$         & ...        & $35.6_{-1.0}^{+0.1}$      &  34.0/32 \\
\hline 
\end{tabular}
\end{table*}
 
\section{Discussion}

Fig. \ref{fig:transition} clearly shows that the flux below 7 keV decreases significantly in the transition.
The optical emission is also tightly correlated with the soft X-ray flux (and is anticorrelated with the hard X-ray flux), which suggests that at least a fraction of the optical emission could be produced either in the inner hot corona through the synchrotron self-Compton mechanism (similarly to Swift J1753.5--0127; \citealt{KVT16}), or that the rapid spectral hardening reduces the amount of ionizing UV and soft X-ray photons, thus leading to less significant optical reprocessing in the outer disk.
These observational facts allow us to exclude the possibility that the observed transition is related to the uncovering of some previously obscured region of the accretion flow because this scenario cannot explain the soft X-ray and optical flux reduction in the transition.
Therefore, we are left with two possibilities: either the event is a classical spectral state transition that occurred while the source was obscured by the outflow, or something must have changed in the disk outflow itself.
The interpretation, however, depends on the spectral model used.

The reflection model fits the data significantly worse than the dual absorber model.
However, the high spectral resolution \emph{NuSTAR} data taken on 2015 June 25 required two reflecting media to explain the iron line profile \citep{WMT17}. 
Moreover, the high-resolution \emph{Chandra} spectra showed that several iron-line species contributed to the emission, and that they occasionally showed P~Cyg profiles \citep{King2015}, which we obviously cannot detect in our low-resolution JEM-X data.
Thus, the poor fits in the low state and the post-transition could be related to overly simplistic assumptions and therefore should not be discarded. 

In this reflection scenario, the pre-transion spectrum is well modeled with a strong disk component, which then disappears in the post-transition spectrum.
It might therefore be tempting to interpret the transition as an evaporation event or as a discrete ejection of the inner accretion disk.
In this case, the change in photon index from $\Gamma \approx 2.9$ to $1.9$ would be analogous to the HSS-LHS (or VHS-LHS) spectral transition seen in many other BHBs \citep{DGK07}.
However, in this scenario it is hard to see how the primary coronal emission can be entirely hidden in the pre-transition spectrum, such that we only see its reflection component, while at the same time we see the cool, geometrically thin disk that extends down to the ISCO.

The interpretation of the dual-absorber model fits is much more straightforward because the parameters of the absorbers do not vary at all. 
In all three spectra we have a fully covering column with $N_{\rm H}\approx 2 \times 10^{23}$ cm$^{-2}$, and a partial covering column with $N_{\rm H,\,pcf} \approx 5 \times 10^{24}$ cm$^{-2}$ and $\textrm{PCF}\approx0.9$.
Here, the transition is simply caused by a photon index variation from $\Gamma \approx 3.7$ to $2.3$.
The changes of these parameters correspond to a transition from a USS, as seen in Cyg~X--3 and GRO~J1655--40 \citep{SZMc08,HZS09,KHMcC10,UKW15},  to a more typical VHS spectrum. 

There are a few examples of state transitions that can also be significantly faster than the typical case of a few days (e.g., \citealt{KTM11, KDT13, DMC14}).
For example, \citet{TKK99} found from \emph{RXTE} and \emph{OSSE} data that the spectrum of GRO~J1655--40 changed from a typical soft-state spectrum to a VHS spectrum within roughly two hours.
GRS~1915+105 is also known for its remarkably complex variability and fast spectral transitions taking place on timescales of seconds \citep{BMK97}. 
It can also display a soft $\Gamma \approx 3$ power-law spectrum that shows no cutoff up to 600~keV \citep{ZGP01}, much like V404~Cyg in several epochs \citep{Sanchez-Fernandez2017}.
However, these variability patterns tend to be repetitive and/or cyclical on relatively short timescales, while we here observe a rapid transition that is sustained for a long period (see fig. 13 of \citealt{AGSF18}, for one quasi-similar case). 
For this intrinsic spectral transition scenario to work for V404 Cyg data, the mechanism responsible for it must therefore be unique to it.

It is not certain, however, that we do see a bona fide spectral transition in the traditional sense because the timescales involved are so rapid. 
In the hot-flow paradigm, the thin disk is believed to be truncated at a large radius from the ISCO of the BH in the hard state (see \citealt{DGK07,PV14}, for review) as well as in the very high state \citep{KD04,TKY12,HUS14,KD16}.
In both cases, the inner disk is thought to puff up to a geometrically thick, but optically thin inner hot flow.
In this paradigm, the (V/L)HS-to-HSS transitions are considered to originate from a condensation of this hot, puffed-up disk into a geometrically thin one \citep{MLM09}, while the reverse HSS-to-LHS transition is instead attributed to an evaporation of the geometrically thin disk back to a thick one \citep{MLM00}.
Depending on the fitting model, the V404~Cyg transition could either be analogous to a VHS-LHS transition (reflection case) or to an USS-VHS transition (two absorber case).
In the latter case, which fits the data much better, there would need to be a disk evaporation event in 15~s.
This timescale is on the same order than the viscous timescale of the inner disk around a 10 solar mass BH with a modest scale height ($t_{\rm visc} =  R^{3/2}\alpha^{-1}(H/R)^{-2}(GM)^{-1/2}$; see, e.g., \citealt{Pringle1981}); that is, by assuming $\alpha = 0.1$, $H/R = 0.1$ and $M = 9\,M_\odot,$ we find $t_{\rm visc} \sim 4$ s at a radius of 256 km (which corresponds to 10 Schwarzschild radii, where most of the X-ray emission is produced).
However, in the coronal models of \citet{MP07}, the disk evaporation or condensation occurs on a timescale of about one day, and the fastest changes that could be related to the observations here also take place in matter of hours, not seconds. 
Moreover, a state transition from an HSS spectrum with a dominating disk component and a power-law tail of $\Gamma\sim2.2$ to a power-law-dominated $\Gamma\sim1.7$ spectrum is typically observed when a source is near the Eddington limit \citep{Maccarone2003,VMKM2019}, whereas here the \emph{\textup{observed}} luminosity in this V404 Cyg transition is lower than 1\%\ of the Eddington limit. 
In addition, soft to hard (or hard to soft) transitions in BH transients typically take place on timescales anywhere between 4--10 days in some cases \citep{KDT13} and up to $\sim20$~days in others \citep{KTM11, DMC14}. This range of state transition timescales was theoretically predicted already in the 1970s \citep{Ichimaru1977}.

Another source, V4641~Sgr (also known as SAX J1819.3--2525), which is a microquasar that also seems to be embedded in a thick envelope during its X-ray outbursts \citep{Revnivtsev2002}, is analogous to V404~Cyg in many respects.
Just like in V404~Cyg, the local absorption column of V4641~Sgr is highly variable on timescales of a few tens of seconds \citep{Revnivtsev2002,MB06}, and it has similar disk-wind signatures in the optical lines \citep{MDTG18}.
Moreover, an analogous counterexample of rapid spectral variability was observed in \emph{RXTE} data on 1999 September 15.
A typical hard-state spectrum evolved to a soft power-law spectrum. The flux was a factor of ten lower within a few minutes (see fig.~7 of \citealt{Revnivtsev2002}).
This similarity suggests an alternative explanation for the observed in V404~Cyg transition: the changes occur in the disk wind rather than in the intrinsic spectrum.
A possible mechanism for the observed spectral transition in this scenario would be a density drop of the Compton-thick disk wind that significantly reduces the amount of Compton down-scattering in the wind (and simultaneously causes less reflection away from our line of sight), much as has been discussed for Cyg X--3 by \citet{ZMG10}.\footnote{The best-fitting parameters for the absorber are constant in the transition. The speculated wind density drop would then have to mean that in the low state, the partially covering column would have to be much higher than the best-fitting value, i.e.,  initially optical depth  drops from $\tau \longrightarrow \infty$ to $\tau \sim 1$ in the transition, which would correspond to the post-transition value of $N_{\rm H,\,pcf} \approx 5 \times 10^{24}$ cm$^{-2}$.
This could well be the case because our absorber model does not take into account the scattering of the radiation away from our line of sight when $\tau \gtrsim 1$.}

In the down-scattering scenario the observed spectrum would therefore only appear much softer in the low-flux state than the incident one  (see, e.g., \citealt{GHM94,MY09}, for analogous cases in AGN tori).
Here too, however, we run into difficulties with the fast 15 s timescales involved.
That the low-flux state before and the higher flux state after the transition both seem rather steady for more than an hour suggests that this is not simply some discrete disk-wind clump that conveniently moves away from our line of sight to reduce the wind density.
Rather, the entire disk-wind structure should change, but this should occur on a timescale of $t_\textrm{wind} \equiv R_\textrm{wind} / v_\textrm{wind} \sim 100 - 10000$ s when we assume that the wind is indeed launched either from a distance $R_\textrm{wind} \sim 10^{10}$ cm \citep{MDCMS16} or from $R_\textrm{wind} \sim 10^{12}$ cm (where the X-ray emission lines seem to be produced, \citealt{King2015}) with the observed velocity of about $v_\textrm{wind} \sim 1000\, \textrm{km s}^{-1}$ \citep{King2015, MDCMS16}.
Even in the most favorable case of $v_\textrm{wind} \approx 3000\, \textrm{km s}^{-1}$ and $R_\textrm{wind} \approx 1.5 \times 10^{10}$ cm \citep{MDCMS16}, the timescale is still about $t_\textrm{wind} \approx 50$ s, which is a factor of a few longer than the observed spectral transition timescale.
To match the timescales, the outflowing material that produces the Compton-thick down-scattering cloud is therefore required to be closer to the BH than the inferred launching radius of the thermal winds.

Other speculative possibilities are that i) the transition could be due to some dynamic or thermal timescale instability in the disk itself (e.g., \citealt{Jiang_2013}). 
Additionally, ii) the rapid transition could be related for example to the inner disk precession and discrete ejections, which is inferred from the radio VLBI observations to change the inner disk (and hence jet) orientation on timescales of minutes \citep{MJTS19}.
Even more speculatively, iii) a drop in the disk scale height from a geometrically thick ($H/R\sim 1$) to a geometrically thinner disk with $H/R \sim 0.1$ could also result in the observed transition.
Because the inclination of V404 Cyg with respect to our line of sight is measured to be about 67$^{\circ}$ \citep{Khargharia2010}, in this hypothetical transition the thick disk itself would no longer obscure the innermost part of the accretion flow, leading to the rapid flux jump and spectral hardening.
But again, in all of these speculative cases is not obvious how the disk height could change this quickly over a large range of radii to accommodate such scenario.

In summary, the rapid spectral transition observed from V404 Cyg poses an interesting case to study the dynamics of the innermost accretion inflow and the disk winds or outflows.
It is not clear, however, which physical process is responsible for the 15 s timescale transition. 
Classical spectral state transitions typically occur on much longer timescales, as do other mechanisms. 
The reduction of Compton down-scattering in the inner disk seems like the simplest scenario that can accommodate the rapid 15 s transition, but alternatives cannot be ruled out.

\begin{acknowledgements}
The authors thank the anonymous referee for useful comments that helped to improve the manuscript.
JJEK acknowledges support from the Spanish MINECO grant ESP2017-86582-C4-1-R and from the Academy of Finland grant 295114.
JAG is grateful for the support from the Spanish MINECO grant ESP2017-87676-C5-1-R.
SEM acknowledges support from the Violette and Samuel Glasstone Research Fellowship programme and from the UK Science and Technology Facilities Council (STFC).
AV acknowledges support from the Academy of Finland grant 309308, and the Ministry of Science and Higher Education of the Russian Federation grant 14.W03.31.0021.
Based on observations with \textit{INTEGRAL}, an ESA project with instruments and science data centre funded by ESA member states (especially the PI countries: Denmark, France, Germany, Italy, Switzerland, Spain), and Poland, and with the participation of Russia and the USA.
\end{acknowledgements}

\bibliographystyle{aa} 

\end{document}